\RequirePackage{ifpdf}
\ifpdf 
\documentclass[pdftex]{sigma}
\else
\documentclass{sigma}
\fi

\begin{document}
\allowdisplaybreaks

\renewcommand{\PaperNumber}{061}

\FirstPageHeading

\ShortArticleName{Constructing Soliton and Kink Solutions of PDE
Models in Transport and Biology}

\ArticleName{Constructing Soliton and Kink Solutions\\ of PDE
Models in Transport and Biology}

\Author{Vsevolod A. VLADIMIROV, Ekaterina V. KUTAFINA and Anna
PUDE\L KO}

\AuthorNameForHeading{V.A. Vladimirov, E.V. Kutaf\/ina and A.
Pude\l ko}

\Address{Faculty of Applied Mathematics  AGH University of Science and Technology,\\
Al. Mickiewicza 30, 30-059 Krak\'{o}w, Poland}
\Email{\href{mailto:vladimir@mat.agh.edu.pl}{vladimir@mat.agh.edu.pl},
\href{mailto:vsan@rambler.ru}{vsan@rambler.ru}}

\ArticleDates{Received November 30, 2005, in f\/inal form May 24,
2006; Published online June 19, 2006}

\Abstract{We present a review of our recent works directed towards
 discovery of a~periodic, kink-like and soliton-like travelling
wave solutions within the models of transport phenomena and the
mathematical biology.  Analytical description of these wave
patterns is carried out by means of our modif\/ication of the
direct algebraic balance method. In the case when the analytical
description fails, we propose to approximate  invariant travelling
wave solutions by means of an inf\/inite series of exponential
functions. The ef\/fectiveness of the method of approximation is
demonstrated on a hyperbolic modif\/ication of Burgers equation.}


\Classification{35C99; 34C60; 74J35}

\section{Introduction}

It is well known that  methods of obtaining  general solutions for
the majority of nonlinear evolutionary PDEs actually do not
exist. The already known methods of integration, such  as the
inverse scattering technique, are applied merely to so called
completely integrable equations~\cite{dodd}. These equations form
a very important but relatively small family, which does not
include e.g.\ nonlinear dissipative models, describing the
spatio-temporal patterns formation and evolution~\cite{prigogine}.
One of the few alternatives  to  numerical methods of the
investigations of those models that are not completely integrable
are the symmetry methods \cite{ovs,olver} which are ef\/fective
when the system under consideration admits a non-trivial group of
transformations. Yet, these methods cannot be treated as universal
tools for obtaining analytical solutions. In fact, the
self-similar reduction, when applied to a nonlinear PDE, gives
another nonlinear dif\/ferential equation with a fewer number of
independent variables and it is dif\/f\/icult to expect that the
procedure will lead to quadratures in every particular case.
Obviously the chances to obtain this way the analytical
description of solutions with the required properties and
asymptotic behavior (such as e.g.~pe\-ri\-o\-dic, kink-like or
so\-li\-ton-like tra\-vel\-ling wave so\-lu\-ti\-ons) are,
ge\-ne\-rally speaking, very small.

In the case when the self-similar reduction leads to a system of
autonomous ODEs the answer to the question on whether the
solutions with the given properties do exist can be obtained by
means of qualitative theory methods~\cite{andron,G-H}, which allow
for  studying of whole families of solutions with the given
symmetry. The qualitative theory methods are very ef\/fective if
one wants to know whether the family of  self-similar solutions
possesses solutions with the given properties. Yet in many cases
mere information about their existence is not suf\/f\/icient. Very
often one needs to have an analytical expression for the given
solution and this is the case when it is necessary to investigate
the inf\/luence of  small addends upon the stability and the
evolution properties of the invariant solutions.

Compared to classical symmetry methods,  more powerful tools for
f\/inding out solutions with the given properties, can be provided
by so-called direct algebraic balance method, based upon
representing solutions in the form of rational algebraic
combinations of properly chosen functions, closed with respect to
the algebraic operations and dif\/ferentiation
\cite{fan,nikbar,baryur,vladku}. However, within this method one
must solve, as a rule, very large systems of nonlinear algebraic
equations. Regardless of this purely technical reason, the
ansatz-based method cannot be treated as the remedy, because there
are known evolution PDEs which possess analytical solutions that
cannot be obtained, at least, within a particular set of
elementary functions (hyperbolic, trigonometric, polynomial,
etc.).  And it is rather doubtful that a fully universal ansatz
for f\/inding out the analytical solutions to nonlinear PDEs can
ever be found. Nevertheless, it is  possible to answer the
question of whether the solution of this or that sort does exist
within the given set of invariant solutions, whenever the
procedure of a self-similar reduction leads to the system of
autonomous ODEs. And in the situation when  the direct algebraic
balance method, based on the representing solution we are looking
for in the form of f\/inite combination of the given functions
fails, we propose to express it by means of the inf\/inite series.

The aim of this study is to review our recent work  directed
towards the description of the  already mentioned wave patterns,
i.e.\ the kink-like and so\-li\-ton-like tra\-vel\-ling wave
so\-lu\-ti\-ons, in various non-local models, simulating the
transport phenomena. In our earlier work we investigated merely
the set of travelling wave  (TW) solutions, being invariant with
respect to the group of the spatio-temporal translations,
nevertheless,  systems obtained via the symmetry reduction prove
to be suf\/f\/iciently dif\/f\/icult to analyze and  solve
exactly. Therefore we use a complex approach including a
qualitative analysis, our own interpretation of the ansatz-based
method and approximation of the invariant solutions by means of
inf\/inite series of exponential functions.

The structure of the article is as follows.  In Section~2 we
present a certain modif\/ication of the direct algebraic method,
put forward in papers \cite{fan,nikbar,baryur}. Employing this
modif\/ication we obtain the kink-like and soliton-like TW
solutions for the hyperbolic modif\/ication of Burgers equation,
telegraph equation, nonlinear d'Alembert equation  and the
somitogenesis model. In Section~3 we present the results of a
qualitative investigations of a set of TW solutions for the
hyperbolic modif\/ication of Burgers equation. It demonstrates the
existence of the TW solution-like solutions, yet these solutions
cannot be described by the f\/inite algebraic combination of
elementary functions. Therefore we propose to approximate the
invariant solitary wave solutions by an inf\/inite series of the
exponential functions.

\section{Exact soliton-like and kink-like solutions\\
for transport equations}

\subsection[Modification of the direct algebraic method and its applications]{Modif\/ication
of the direct algebraic method and its applications}

The essence of the  {\it  unified direct algebraic method} in
E.~Fan's paper~\cite{fan}, is based on the observation that
particular solutions of any system of PDE's
\begin{gather}\label{genevol}
H_\nu\big(u^i,u^i_t, u^i_x,u^i_{xx},\dots \big)=0,\qquad
i=1,2,\dots, m, \quad \nu =1,2,\ldots,n
\end{gather}
which do not depend on $(t,x)$ coordinates in explicit form,  can
be presented as a polynomial expansion
\begin{gather}\label{genansatz}
u^i(\xi)=\sum_{\mu=0}^{s}a_\mu^i \phi^\mu(\xi), \qquad \xi=x+vt,
\end{gather}
where $a_\mu$ are unknown parameters, while function $\phi(\xi)$
satisf\/ies the equation
\begin{gather}\label{phieqn}
\dot\phi(\xi)=\pm \sqrt{\sum_{\nu=0}^r c_\nu \phi^\nu(\xi)}.
\end{gather}
Note that in order to obtain non-trivial solutions, one should
``balance'' the numbers $r$ and $s$ (for  details see \cite{fan}).

Depending on the conditions posed on  the parameters $c_\nu,$
solutions to the equation (\ref{phieqn}) are expressed by elliptic
Jacobi functions, and in particular cases by hyperbolic or
trigonometric functions \cite{fan}. The properties of these
functions are inherited by the solution of the initial system
which can be presented in the form (\ref{genansatz}). In fact this
methodology is constructive and algorithmic  if the functions
$H_{\nu}\left(\cdot,\cdot,\dots  \right)$ arising in
(\ref{genevol}) are algebraic or rational ones. With this
assumption, for any set of functions $\left\{u^i \right\}_{i=1}^m$
that can be  represented as (\ref{genansatz}), the LHS of the
equations (\ref{genevol}) can be presented in the form of f\/inite
linear combinations of functions $\phi^k(\xi)$ and $\phi^l(\xi)
\dot\phi(\xi)$,  $k,l=0,1,2,\dots$:
\[
H_\nu\big(u^i,u_t^i,u_x^i,\ldots\big)=\sum_{\alpha=0}^j
b_{\nu\alpha}\phi^\alpha (\xi) + \dot\phi(\xi) \sum_{\gamma=0}^s
h_{\nu\gamma}\phi^\gamma (\xi), \qquad \nu=1,\ldots,n.
\]
 When equating to zero all the coef\/f\/icients
of these decompositions, one obtains a system of nonlinear
algebraic equations, that determine the particular solutions of
the initial system of PDE's (for more details see \cite{fan}). It
must be stressed that the method does not guarantee  obtaining of
the general solution, even within the set (2), for its
ef\/fectiveness depends on one's ability to solve a system of
nonlinear algebraic equations. On the other hand, it enables
calculation of particular solutions possessing some predictable
properties.

An enforced version of Fan's method was  put forward recently in
\cite{nikbar,baryur}. It was used for determining of exact
solutions of the equation
\[
u_t+Auu_x-\kappa u_{xx}=f(u).
\]
More precisely, the following ansatz was  proposed:
\begin{gather}\label{ansnik}
u=\left[\frac{z'(\xi)}{z(\xi)}\right]^k, \qquad \xi=x+vt+x_0,
\qquad k=1,2,\ldots,
\end{gather}
where $z(\xi)=\sum\limits_{\mu=0}^k a_\mu \phi^\mu(\xi)$, and
$\phi(\xi)$ is a function satisfying equation~(\ref{phieqn}). Due
to this combination, the multi-parameter families of the exact
solutions were obtained in the situation when the pure form of
Fan's methodology does not work \cite{nikbar}.

On analyzing dif\/ferent versions of the ansatz-based method, one
can conclude that ef\/fec\-ti\-ve\-ness of their employment is
based on the mere fact that the family of functions
$\phi^\mu(\xi)$ and $\phi^\nu\dot{\phi }(\xi)$ is closed with
respect to the algebraic operations and dif\/ferentiation. In view
of this, a~quite natural generalization of the already mentioned
ansatzes can be as follows:
\begin{gather}\label{vkgenans}
 u=\frac{f(\xi)}{g(\xi)}=\frac{\sum\limits_{\mu=0}^{m_1} a_\mu \phi^\mu(\xi)+
 \dot\phi(\xi)\sum\limits_{\nu=0}^{m_2} b_\nu \phi^\nu(\xi)}
 {\sum\limits_{\lambda=0}^{n_1} c_\lambda \phi^\lambda(\xi)+
 \dot\phi(\xi)\sum\limits_{\kappa=0}^{n_2} d_\kappa \phi^\kappa(\xi)},
 \end{gather}
where  the function $\phi(\xi)$   still satisf\/ies the equation
(\ref{phieqn}), but, in contrast to (\ref{ansnik}), dependence
between the functions $f$ and $g$ is not assumed from the very
beginning. Ef\/fectiveness of the ansatz (\ref{vkgenans}) is
demonstrated by applying it to some transport equations given
below.

\subsection{Soliton-like and kink-like TW solutions\\ for the generalized transport equations}

Let us consider the following equations:
\begin{gather}\label{vkhypeq}
\tau u_{tt}+Auu_x+Bu_t-\kappa u_{xx}=f(u)=\sum_{\nu \in
I}\lambda_\nu u^\nu,
\end{gather}
where $\tau$, $A$, $B$, $\kappa$ are non-negative constants.
 For $A=0$ equation (\ref{vkhypeq}) coincides with the
nonlinear telegraph equation; for $A\ne 0$ it coincides with the
hyperbolic generalization of  Burgers equation, while for $A=B=0$
-- with the nonlinear d'Alembert equation. The hyperbolic
modif\/ications of nonlinear transport equations arise naturally
when the ef\/fects of temporal non-locality are taken into account
\cite{makar95}.

Assuming that the solitons and kinks can be expressed by powers of
function ${\rm sech}(\xi)$, which is the particular solution of
equation (\ref{phieqn}) and function $\sinh(\xi)$ which appears in
the odd derivatives of the function ${\rm sech}(\xi)$, we use the
following ansatz:
\begin{gather}\label{shans}
u(\xi)=\frac{f(\xi)}{g(\xi)}=\frac{\sum\limits_{\mu=0}^{m_1}a_\mu
{\rm sech}^\mu (\alpha\xi)+
\sinh(\alpha\xi)\sum\limits_{\nu=0}^{m_2}b_\nu {\rm sech}^\nu
(\alpha\xi)} {\sum\limits_{\gamma=0}^{n_1}c_\gamma {\rm
sech}^\gamma (\alpha\xi)+\sinh(\alpha\xi)
\sum\limits_{\sigma=0}^{n_2}d_\sigma {\rm sech}^\sigma
(\alpha\xi)}
\end{gather}
or, equivalently,
\begin{gather}\label{expans}
u(\xi)=\frac{\sum\limits_{\mu=0}^{m}a_\mu \exp(\mu\alpha\xi)}
{\sum\limits_{\nu=0}^{n}b_\nu \exp(\nu\alpha\xi)}.
\end{gather}

Inserting  the ansatz (\ref{shans}) ( or (\ref{expans})) into
(\ref{vkhypeq}) and executing all the necessary operations, we
obtain an algebraic equation containing, respectively, functions
${\rm sech}^\mu(\alpha\xi)$, ${\rm
sech}^\nu(\alpha\xi)\sinh(\alpha\xi)$ or $\exp{[\mu \alpha \xi]}$.
Deeming them functionally independent and equating to zero the
corresponding coef\/f\/icients, we go to the system of algebraic
equations. We do not demonstrate the details of these calculations
since they are simple but cumbersome. The calculations were
performed through the aid of a~software application
``Mathematica''. The results obtained are presented below.

I. For arbitrary $A$, $B$ and $f(u)=\lambda_0+\lambda_1
u(t,x)+\lambda_2 u(t,x)^2+\lambda_3 u(t,x)^3$ the function
\begin{gather}\label{vk1_1}
u(t,x)=\frac{a_0+a_1e^{ \alpha (x+vt)}}{b_0+b_1e^{ \alpha (x+vt)}}
\end{gather}
satisf\/ies equation (\ref{vkhypeq}) if the following conditions
hold:
\begin{gather}
\lambda_0=\frac{-a_0a_1\alpha}{ \Delta^2 }(Bv\Delta+h\Theta),
\quad \lambda_1=\frac{1}{b_0b_1\Delta^2}\left[\alpha
b_0b_1(Bv\Theta\Delta+h\Theta^2)+\lambda_3a_0a_1\Delta^2\right],
\nonumber\\
\lambda_2=\frac{-1}{b_0b_1\Delta^2}\left[\alpha
b_0^2b_1^2(Bv\Delta+h\Theta)+\lambda_3\Delta^2\Theta\right], \quad
A=\frac{-1}{\alpha b_0b_1\Delta}\left[-2h\alpha
b_0^2b_1^2+\lambda_3\Delta^2\right].\label{vk1_1aux}
\end{gather}
Here and henceforth we use the notation $
h=\alpha(v^2\tau-\kappa)$, $\Delta=a_1b_0-a_0b_1$, $
\Theta=a_1b_0+a_0b_1.$ Equation (\ref{vk1_1}) def\/ines a
kink-like regime when $b_0 b_2>0$ and
${a_0}/{b_0}\neq{a_2}/{b_2}$. Using the
conditions~(\ref{vk1_1aux}), we can express the unknown parameters
from formula~(\ref{vk1_1}) by the parameters characterizing
equation~(\ref{vkhypeq}), yet, in the general case it is too
cumbersome. It is much easier to do when
$\alpha=2\sqrt{-\lambda_0\lambda_2}/v$,
$a_0=-a_1=\sqrt{-\lambda_0/\lambda_2}$, $b_0=b_1=1$. From these
conditions we obtain the solution:
\[
u(t, x)=\sqrt{-\lambda_0/\lambda_2}
\tanh\left[\frac{\sqrt{-\lambda_0\lambda_2}}{v}(x+vt)\right],
\]
where
\[
 v=\frac{\lambda_2\big(A B+\sqrt{A^2 B^2-8\kappa \lambda_3+16\kappa \lambda_2^2 \tau}\big)}
 {2\lambda_3-4\lambda_2^2 \tau}, \qquad \lambda_1=\frac{\lambda_0 \lambda_3}{\lambda_2}.
\]
For  $\kappa=1$, $\tau=0$, $A=0$, $B=1$ function (\ref{vk1_1})
coincides with the solution obtained in \cite{nikbar}.

Another example of the kink-like solution def\/ined by the
formulae (\ref{vk1_1}), (\ref{vk1_1aux}) is as follows:
\[
u(t,x)=\frac{2}{b_0[1+\exp({2 \alpha \xi})]}, \qquad \xi=x+vt,
\]
where $ b_0={\big(-\lambda_2\pm\sqrt{\lambda_2^2-4 \lambda_1
\lambda_3}\big)}/{\lambda_1}$, $A=\lambda_0=0$,
$\alpha=-{(2\lambda_2+3 b_0 \lambda_1)}/{[4B v b_0]}$, and
\[
v=\frac{\pm\sqrt{\kappa}(2\lambda_2+3 b_0
\lambda_1)}{\sqrt{4\lambda_2^2\tau+4b_0\lambda_2(B^2+3\lambda_1\tau)+b_0^2\lambda_1(2B^2+9\lambda_1\tau)}}.
\]

II. For $A=0$, arbitrary $B$  and
$f(u)=\lambda_0+\lambda_{1/2}u(t,x)^{\frac{1}{2}}+\lambda_1
u(t,x)+\lambda_{3/2} u(t,x)^{\frac{3}{2}}+\lambda_2 u(t,x)^2$
function
\begin{gather*}
u(t,x)=\left[\frac{a_0+a_1e^{ \alpha (x+vt)}}{b_0+b_1e^{ \alpha
(x+vt)}} \right]^2
\end{gather*}
satisf\/ies  the equation~(\ref{vkhypeq}) providing that the
following conditions hold:
\begin{gather*}
\lambda_0={2a_0^2a_1^2\alpha h}/{\Delta^2}, \qquad
\lambda_{1/2}={-2a_0a_1\alpha}(3h\Theta+Bv\Delta)/{\Delta^2},
\\
\lambda_1={2\alpha}(h(3\Theta^2-\Delta^2)+Bv\Delta\Theta)/{\Delta^2},
\qquad
\lambda_{3/2}={-2b_0b_1\alpha}(5h\Theta+Bv\Delta)/{\Delta^2},
\\
\lambda_2={6b_0^2b_1^2h\alpha}/{\Delta^2}.
 \end{gather*}
 This solution def\/ines the solitary wave regime if $b_0 b_1>0 $,
${|a_0|}/{|b_0|}={|a_2|}/{|b_2|} $ while ${a_0}/{b_0}\neq
{a_2}/{b_2}.$

III. For $B=0$, arbitrary $A$  and $f(u)= \lambda_1
u(t,x)+\lambda_3 u(t,x)$ function
\begin{gather}\label{vk3}
u(t,x)=\frac{a_1e^{\alpha \xi}+a_2e^{2\alpha \xi}}{-a_1^3-3 a_1^2
a_2 e^{\alpha \xi}+3 a_1 a_2^2 e^{2\alpha \xi}+a_3^3 e^{3\alpha
\xi}}, \qquad \xi=x+vt,
\end{gather}
satisf\/ies (\ref{vkhypeq}), when $\lambda_1$, $\lambda_3$ are
positive and the parameters are as follows: $ 6 a_1
a_2=-\sqrt{\lambda_3/\lambda_1}$, $\alpha={\sqrt{\lambda_1
\lambda_3}}/{A}$, $
v=\pm\sqrt{\left(A^2/\lambda_3+\kappa\right)/{\tau}}.$ This
solution is always singular, because for arbitrary values of the
parameters the expression in the denominator of the formula
(\ref{vk3}) vanishes  for some  $\xi\in {\mathbb R}$.

IV. Now let us consider the case ${A=B=0.}$

IVa. For $f(u)=\lambda_0+\lambda_1  u(t,x)+\lambda_2
 u(t,x)^2+\lambda_3  u(t,x)^3$
 function
\begin{gather}\label{vk4_1}
u(t,x)=\frac{a_0+2a_1 e^{\alpha (x+vt)}+a_0 e^{2 \alpha
(x+vt)}}{b_0+2 b_1 e^{\alpha (x+vt)}+b_0 e^{2\alpha (x+vt)}},
\end{gather}
satisf\/ies equation (\ref{vkhypeq}) when the following conditions
hold:
\begin{gather}
\lambda_0=a_0 (2a_0^2 b_0-a_1^2 b_0-a_0 a_1 b_1 )\alpha h/\Delta^2,\nonumber\\
\lambda_1=((a_1^2 b_0^2+4a_0 a_1 b_0 b_1+a_0^2 (-6 b_0^2+b_1^2))\alpha h/\Delta^2,\nonumber\\
\lambda_2=3 b_0 (2 a_0 b_0^2-a_1 b_0 b_1-a_0 b_1^2)\alpha h/\Delta^2,\qquad
\lambda_3=-2 b_0 (b_0^2-b_1^2)\alpha h/\Delta^2.\label{vk4_1aux}
\end{gather}
For $a_0\neq 0$, $b_0 \neq 0$, $|a_1|+|b_1|\neq 0$ equation
(\ref{vk4_1}) defines  the soliton-like solution. One of the
parameters contained in (\ref{vk4_1}) can be chosen arbitrarily,
while the rest can be expressed, using~(\ref{vk4_1aux}), as the
functions of the parameters, defining equation (\ref{vkhypeq}). We
omit doing this in the general case, but present one particular
example. Thus, for $b_1=0,  b_0=1$ and  arbitrary $\alpha$ we have
the solution
\[
u(t, x)=\frac{a_0+2a_1e^{\alpha (x+vt)}+a_0e^{2\alpha
(x+vt)}}{1+e^{2\alpha (x+vt)}},
\]
with \[    a_0=-{\lambda_2}/{(3 \lambda_3)}, \qquad
  a_1=\sqrt{2{(\lambda_2^2-\lambda_1 \lambda_3)}/{\lambda_3^2}},
\qquad
 v=\pm\sqrt{\left[\lambda_1-{\lambda_2^2}/{(3 \lambda_3)}+\kappa \alpha^2
\right]/({\tau \alpha^2})},
\]
and the additional condition
$\lambda_0+\lambda_1a_0+\lambda_2a_0^2+\lambda_3a_0^3=0$.

VIb. For $f(u)=\lambda_{1/2} u(t,x)^{{1}/{2}}+\lambda_1
u(t,x)+\lambda_{3/2} u(t,x)^{{3}/{2}}+\lambda_2 u(t,x)^{2}$ we
obtain a soliton-like solution
\begin{gather*}
u(t,x)=\frac{(e^{\alpha (x+vt)}+1)^4}{(b_0e^{2\alpha
(x+vt)}+(2b_0+4b_1)e^{\alpha (x+vt)}+b_0)^2}
\end{gather*}
with
\begin{gather*}
\lambda_{1/2}=-3 \alpha  h/b_1,\qquad  \lambda_1=(12 b_0+4 b_1)\alpha  h/b_1, \\
\lambda_{3/2}=-(15 b_0^2+10 b_0 b_1)\alpha  h /b_1, \qquad
\lambda_2=(6 b_0^2 b_1+6 b_0^3)\alpha  h/b_1.
\end{gather*}

VIc. For $f(u)=\lambda_0+\lambda_{1/2}
u(t,x)^{\frac{1}{2}}+\lambda_1 u(t,x)+\lambda_{3/2}
u(t,x)^{\frac{3}{2}}$ a localized wave packet
\begin{gather*}
u(t, x)=\frac{(a_0e^{2\alpha (x+vt)}+(2a_0+4a_1)e^{\alpha
(x+vt)}+a_0)^2}{(e^{\alpha (x+vt)}+1)^4}
\end{gather*}
def\/ines a solution of (\ref{vkhypeq}) if the following
conditions hold:
\begin{gather*}
\lambda_0=2a_0^2(a_0+a_1) \alpha  h/a_1,\qquad
 \lambda_{1/2}=(9a_0^2+6a_0a_1) \alpha  h/a_1, \\
\lambda_{1}=(12a_0+4a_1) \alpha  h/a_1,\qquad
 \lambda_{3/2}=-5 \alpha  h/a_1.
 \end{gather*}

VId. For $f(u)=\lambda_1 u(t,x)+\lambda_{3/2}
u(t,x)^{\frac{3}{2}}+\lambda_2u(t,x)^2$ the function
\begin{gather*}
u(t,x)=\frac{4e^{2\alpha (x+vt)}}{(a_0e^{2\alpha
(x+vt)}+2a_1e^{\alpha (x+vt)}+a_0)^2}
\end{gather*}
with
\[ \lambda_1=4 \alpha  h,\qquad
\lambda_{3/2}=-10 a_1\alpha  h,\qquad
 \lambda_2=(6 a_1^2-6 a_0^2)\alpha  h
\]
  def\/ines a soliton-like solution of the
equation (\ref{vkhypeq}).

 VIe. Finally, let us consider the equation
\begin{gather}\label{dalamb}
\tau u_{tt}-\kappa u_{xx}=\lambda_0+\lambda_1
u+\lambda_2u^2+\lambda_3u^3.
\end{gather}
Substituting the ansatz $u=\phi(\xi)$, $\xi=x+v t$ into the
equation (\ref{dalamb}), we obtain, after one integration, the
following ODE:
\begin{gather*}
\frac{d\phi}{d\xi}=\pm \sqrt{c_0+c_1 u+c_2 u^2+c_3 u^3+c_4 u^4},
\end{gather*}
where $c_0$ is an arbitrary constant, $ c_1=2\lambda_0/\delta$,
$c_2=\lambda_1/\delta$, $ c_3=2\lambda_2/(3 \delta)$,
$c_4=\lambda_3/(2 \delta),$ $\delta=\tau v^2-\kappa.$  For this
equation the classif\/ication given in \cite{fan} can be applied:
\begin{itemize}\itemsep=0pt
\item[(a)]
 if $\lambda_2=\lambda_0=0$, then equation (\ref{dalamb}) possesses a soliton-like solution
\begin{gather*}
u(t, x)=\sqrt{-2\lambda_1/\lambda_3}{\rm
sech}\big(\sqrt{\lambda_1/\delta}\xi\big),\qquad\lambda_1>0,\qquad
\lambda_3<0;
\end{gather*}
 \item[(b)] if $\lambda_0=\lambda_2=0$,
 then equation (\ref{dalamb}) possesses a kink-like solution
\begin{gather*}
u(t, x)=\sqrt{-\lambda_1/\lambda_3} \tanh\big[\sqrt{-\lambda_1}/(2
\delta) \xi\big],\qquad\lambda_1<0,\qquad \lambda_3>0;
\end{gather*}
 \item[(c)] if $\lambda_0=\lambda_3=0$, then
equation (\ref{dalamb}) possesses a soliton-like solution
\begin{gather*}
u(t, x)=-\left[3\lambda_1/(2\lambda_2)\right]{\rm sech}^2
\big[\sqrt{\lambda_1/\delta} \xi/2\big],\qquad\lambda_1>0.
\end{gather*}
 \end{itemize}

\begin{figure}
\centerline{\includegraphics[width=7.5cm]{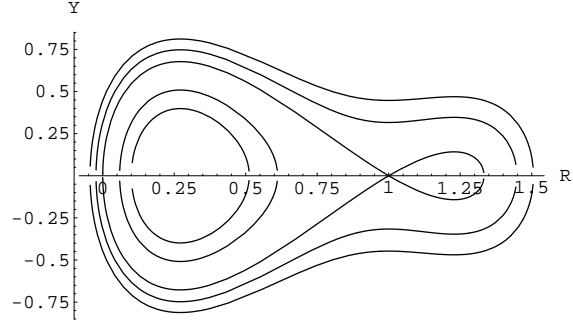}}
\caption{Phase portrait of dynamic system describing the TW
solution of system (\ref{vkhypeq}): case IV.}\label{hamilt}
\end{figure}

More general results can be obtained using the fact that
(\ref{dalamb}) corresponds to a Hhamiltonian case. Using the
standard procedure, we obtain the following integral:
\begin{gather}\label{int}
\xi=\int\frac{du}{\sqrt{2H+{2} {\delta^{-1}}(\lambda_0 u+\lambda_1
u^2/2+ \lambda_2 u^3/3+\lambda_3 u^4/4)}},
\end{gather} where $H$ is
the value of the Hamiltonian function on the given phase
trajectory. The integral~(\ref{int}) can be calculated explicitly
in terms of elliptic functions for dif\/ferent values of $H$ and
as special cases contains all the solutions obtained in part IV.
An example of phase portrait for $\Delta=1,$ $\lambda_0=2,$
$\lambda_1=-11,$ $\lambda_2=15,$ and $\lambda_3=-6$ is presented
in Fig.~\ref{hamilt}.

\subsection{A model of mathematical biology}
Now we consider the following system \cite{daragh}:
\begin{gather}
u_t+Auu_x-u_{xx}=f(u,z),\nonumber\\
z_t=g(u,z),\label{som}
\end{gather}
where $f(u,z)$, $g(u,z)$ are given polynomials, and $A$ is a
positive constant.

Due to the fact that the system is autonomous we apply variables
in a travelling wave form:
\[
u(t,x)=V(\xi),\qquad z(t,x)=U(\xi),
\]
where $\xi=x+\mu t$, and we reduce system (\ref{som}) to the
following system of ODE:
\[
\mu V'+AVV'-V''=f(U,V),\qquad \mu U'=g(U,V).
\]

To simplify the computations, we restrict the family of admissible
polynomials in RHS as follows:
\[
f(U,V)=a_1U+a_2V+a_3UV, \qquad g(U,V)=b_1U+b_2V+b_3UV.
\]
\vskip 0.2 cm

In order to obtain soliton- and kink-like solutions we proposed
the following ansatz
\[
U(\xi)=\frac{c_1e^\xi}{d_0+d_1e^\xi+d_2e^{2\xi}},\qquad
V(\xi)=\frac{p_1e^\xi}{q_0+q_1e^\xi+q_2e^{2\xi}}
\]
that shows the proper asymptotic behavior.

The following kink-like solutions were obtained:
\[
U(\xi)=\frac{p_1}{q_1+q_2e^\xi}, \qquad
V(\xi)=\frac{2q_1}{Aq_1+Aq_2e^\xi},
\]
where
\[
a_1=\mu-1, \qquad a_2=0,\qquad a_3=\frac{q_1(1-\mu)}{p_1},\qquad
b_1=-\frac{Ap_1\mu}{2q_1},\qquad b_2=0,\qquad b_3=\frac{A\mu}{2}.
\]

For other family of parameters
\begin{gather*}
 a_1=\mu^2-1,\qquad
a_2=\frac{-2q_1\mu}{Ap_1},\qquad a_3=\frac{2q_1}{p_1},\\
b_1=-\frac{Ap_1(\mu^2-1)\mu}{2q_1},\qquad b_2=\mu^2, \qquad
b_3=-A\mu
\end{gather*}
the ansatz provides the following soliton-like solutions:
\begin{gather*}
U(\xi)=\frac{4p_1q_2e^\xi}{(q_1+2q_2e^\xi)^2},
\\
V(\xi)=\frac{8q_1q_2e^\xi}{A(q_1+2q_2e^\xi)(2q_2(\mu-1)e^\xi+q_1(\mu-1))}.
\end{gather*}

\section{Approximated soliton-like solutions\\ of  the generalized Burgers equation }

\subsection{Existence of the soliton-like regimes within the set of TW solutions}

 We  showed in the previous section that the ansatz (\ref{expans})
is quite ef\/fective in obtaining  travelling wave solutions with
given properties and asymptotics. Yet recently we realized that it
is quite impossible to describe soliton-like solutions of the
following generalization of the Burgers equation (GBE from now
on):
\begin{gather}\label{genburg}
\tau u_{tt}+uu_x+u_t-\kappa u_{xx}+\gamma u (u^2-z_0^2)=0,
\end{gather}
using the ansatz (\ref{expans}), which is more general than the
generalized Hirota-like ansatzes employed in~\cite{nikbar,baryur}.
The f\/irst thing we are going to do in the situation when the
ansatz-based methods fail, is to perform a qualitative study of
the ODE system describing the TW solutions of the GBE and answer
the question whether (and when) soliton-like solutions exist.

Below we analyze in detail the whole set of TW solutions of
equation (\ref{genburg}). The results of this analysis are used in
the following section in constructing the approximated
soliton-like solutions. Inserting the ansatz $u=U(\xi)=U(x+\mu t)$
into the equations (\ref{genburg}) we obtain the second order ODE
\begin{gather}\label{dissode}
h \frac{d^2 U}{d \xi^2}+\frac{d U}{d \xi} (\mu+U)+\gamma
U\left(U^2-z_0^2\right)=0.
\end{gather}
$h=\tau \mu^2-\kappa$. The scaling transformation
\begin{gather*}
\bar U=\frac{U}{z_0}, \qquad T=\sqrt{\frac{\gamma z_0^2}{h}}\xi,
\end{gather*}
leads us to the equation
\begin{gather}\label{redode}
\frac{d^2 \bar U}{d T^2}+A \frac{d \bar U}{d T} (\bar\mu+\bar
U)+\bar U\left(\bar U^2-1 \right)=0,
\end{gather}
where $A=1/\sqrt{h \gamma^2}$ and $\bar \mu={\mu}/{z_0} $ (we omit
bars over the variables in the following formulae).

Let us rewrite (\ref{redode}) as the  following system of the
f\/irst order ODEs:
\begin{gather}\label{candis}
\frac{d U}{d T}=-W, \qquad \frac{d W}{d T}=U (U^2-1)-A W \left(\mu
+ U \right).
\end{gather}

System (\ref{candis}) has three stationary points: $B_0 (0, 0)$,
$B_1=(-1, 0)$ and $B_2=(1, 0)$. Analysis shows \cite{romp05} that
the critical point $B_0$ is always a saddle. Depending on the
values of the para\-meters, two other critical points are either
nodes or foci. As was declared earlier, we are looking for the
homoclinic trajectory  and since the dynamical system
(\ref{candis}) is of dissipative type, this trajectory can exist
only for special values of the parameters. In our attempts to
capture the homoclinic loop we begin with studying the condition
of the  limit cycle creation in proximity of the critical point
$B_2$. If there exists a limit cycle lying in the right half plane
and surrounding the critical point $B_2$, then its evolution in
the vicinity of the unmovable saddle point $B_0$, caused by the
change of the driven parameter $\mu$, most probably will lead to
the homoclinic bifurcation.

\begin{figure}
\centerline{\includegraphics[width=7cm]{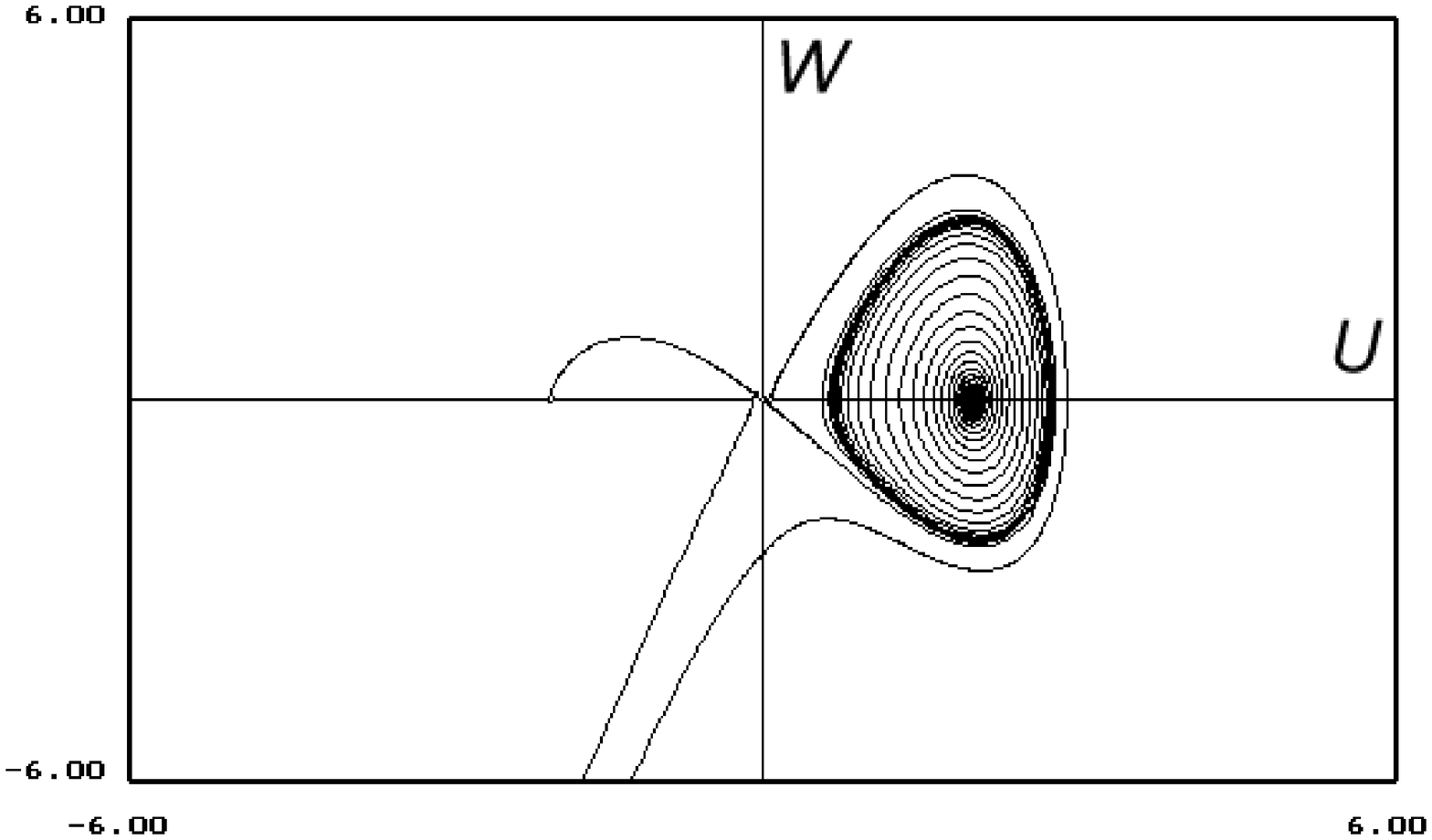}\qquad
\includegraphics[width=7.5cm]{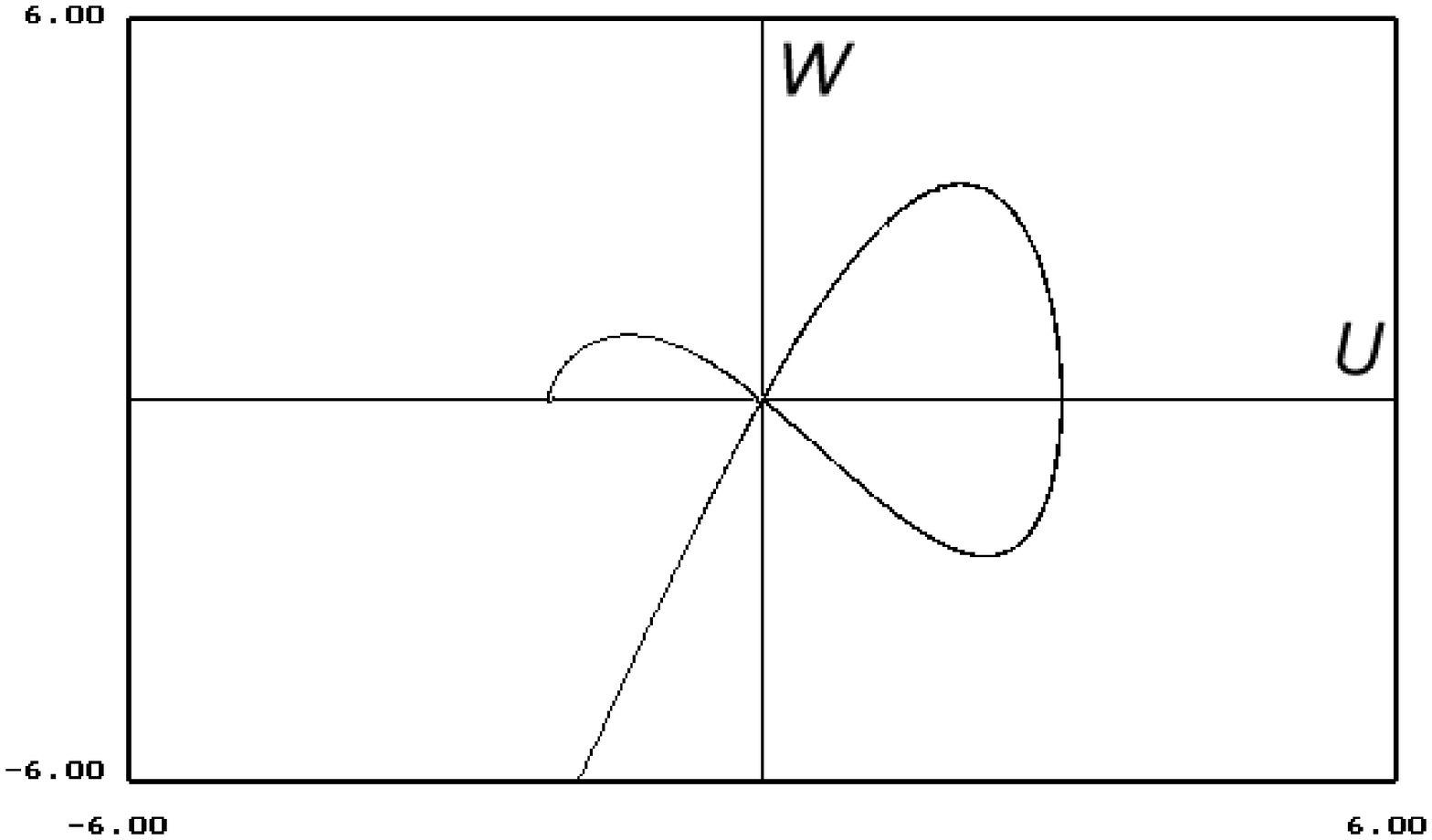}}
\vspace{-1mm}

\hspace*{7.2cm}$a$ \hspace{7.7cm} $b$

\vspace{3mm}

\centerline{\includegraphics[width=7.5cm]{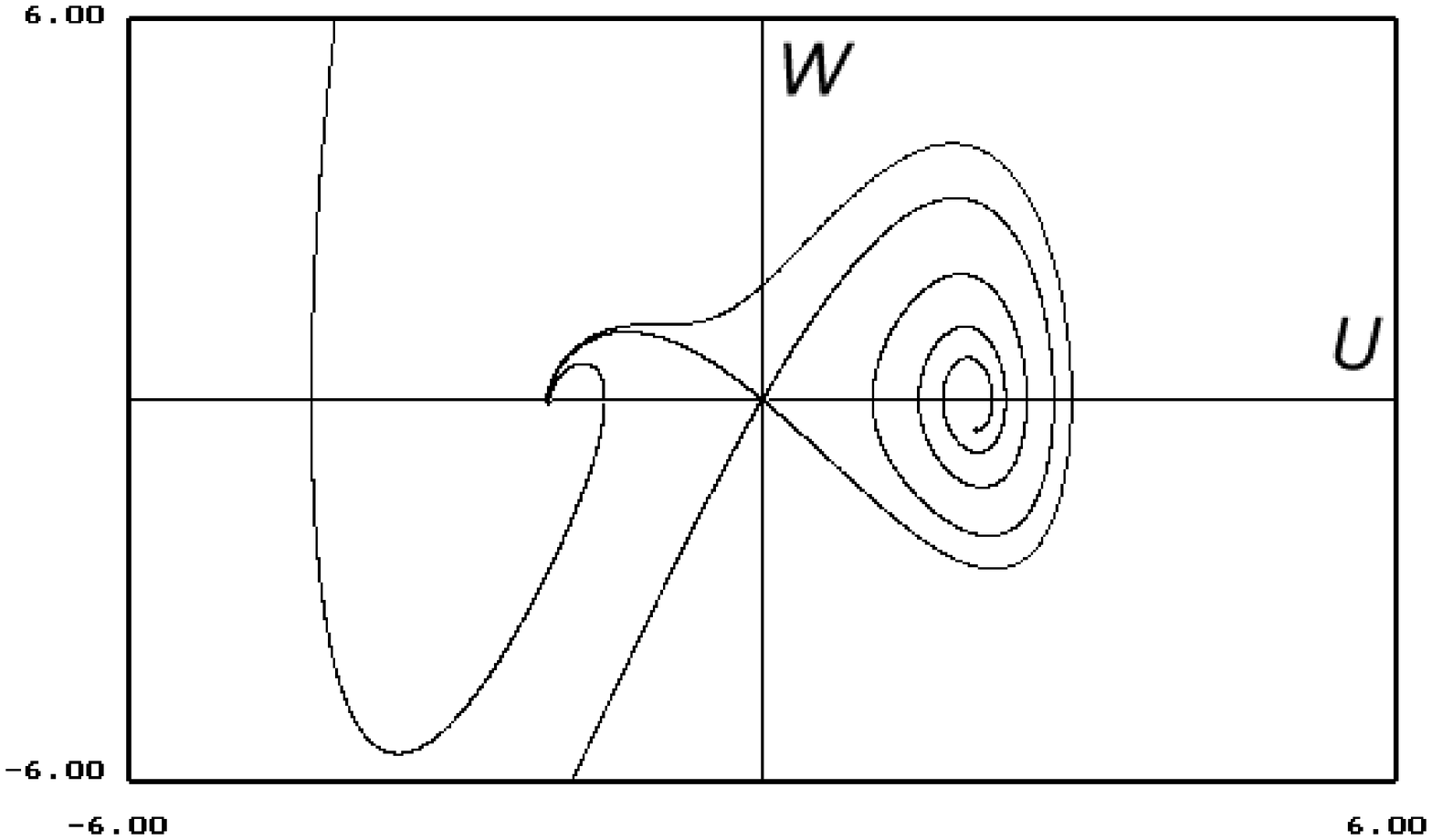}}
\vspace{-1mm}

\hspace*{11.5cm}$c$

\caption{Phase portraits of system (\ref{candis}); (a):
$\mu=-0.9$; (b): $\mu=-0.836$; (c): $\mu=-0.8$.}\label{fig:4-1}
\end{figure}

It can be shown by inspection of the Andronov-Hopf theorem
conditions and their ful\-f\/il\-lment \cite{has,romp05} that in
vicinity of the critical value of the parameter $\mu=-1$ an
unstable limit cycle is created with zero radius. A further
evolution of system (\ref{candis}) was undertaken by means of the
numerical simulation in which we put $A=1$. It shows that the
radius of the periodic trajectory grows with the parameter $\mu$,
until the homoclinic bifurcation takes place at $\mu\approx
-0.836.$ Note that the homoclinic solution intersects the
horizontal axis at the point $U=x_*\approx 1.426095$. On further
growth of the parameter $\mu$ the homoclinic loop is destroyed and
one of the phase trajectories starting at the unstable saddle
$B_1$ reaches the critical point $B_2$. Patterns of the phase
trajectories of system (\ref{candis}) obtained by means of
numerical simulations are shown in Fig.~\ref{fig:4-1}.

Thus, the  homoclinic  loop exists in system (\ref{candis}) merely
at the specif\/ic values of the parameters. The construction of
the approximated soliton-like solution, undertaken below is based
in essential way upon the preliminary information obtained in this
section.

\subsection{Approximation of the soliton-like TW solution to the GBE}

The results of qualitative analysis tell us that equation
(\ref{genburg}) possesses the SW solution for the specif\/ied
values of the parameters. Let us assume that it can be described
by means of  the formula (\ref{expans}).  Using the fact that the
SW solution is represented by the homoclinic trajectory of system
(\ref{redode}), that is bi-asymptotic to the origin, we can use
for its description the following representation
\begin{gather}\label{nogo}
u(\xi)=\frac{\sum\limits_{k=1}^{m}a_k \exp{[k\alpha\xi]}}
{1+\sum\limits_{r=1}^{m+1}b_r \exp{[r\alpha\xi]}},
\end{gather}
that gives us  proper asymptotics when $\xi\to\pm\infty$. The
problem is that the formula (\ref{nogo}) does not describe the SW
solution whenever $m$ is a f\/inite number, as it is stated below.

\begin{lemma} \label{lem:4-1} A solitary wave solution of the equation
\eqref{genburg} cannot be described by the formula~\eqref{nogo}.
\end{lemma}

\begin{proof} Since there is a one-to-one correspondence between the
solutions of equations (\ref{dissode}) and~(\ref{redode}), we
consider the latter one, trying to represent its homoclinic
solution by means of the formula
\begin{gather}\label{nogoscal}
U(T)=\frac{\sum\limits_{k=1}^{m}a_k \exp{[k\alpha T]}}
{1+\sum\limits_{r=1}^{m+1}b_r \exp{[r\alpha T]}}.
\end{gather}
Let us denote the numerator and denominator of the above
expression by $F(T)$, $G(T)$ respectively. Inserting the ansatz
(\ref{nogoscal}) into the equation (\ref{redode}) and multiplying
the resulting equation by $G^3$, we get the expression
\[
G(F'' G-G'' F)+[A(F+\mu G)-2 G'](F'G-G'F)+F(F^2-G^2)=0.
\]
Now, gathering the coef\/f\/icients standing at functions
$\exp{(\alpha T)}$, $\exp{[(3 m+2)\alpha T]}$ and assuming that
the coef\/f\/icients $a_1,$ $a_m$, $b_{m+1}$ are nonzero, we
obtain the following system of algebraic equations:
\[
\alpha^2+A \mu \alpha-1=0, \qquad \alpha^2-A \mu \alpha-1=0,
\]
that leads to a contradiction when $A$ is nonzero.
\end{proof}

Of course, the formula (\ref{nogoscal}) does not give the most
general expression for a solution, having the proper asymptotics
at $T\to\pm\infty$ and belonging to the family (\ref{expans}). For
example, we can take the formula
\begin{gather}\label{moregenans}
U(T)= \frac{\sum\limits_{k=n_1}^{m+n_1}a_k \exp{[k\alpha T]}}
{1+\sum\limits_{r=n_2}^{m+n_1+n_3}b_r \exp{[r\alpha T]}}
\end{gather}
instead of (\ref{nogoscal}) for any  natural numbers $n_1$,
$n_2$,  $n_3$ and $m.$ But our experiments with this formula in
the cases when the numerator and denominator contained,
respectively, two and three terms, gave a result which is
identical to that stated in Lemma~\ref{lem:4-1}. We do not mention
it here because it is too cumbersome. Concerning an expression
containing more terms the result is unclear as yet, but it is
rather doubtful that we would obtain a positive answer by
incorporating extra terms, for a number of extra algebraic
equations accompanying the ``inf\/lation'' of the
formula~(\ref{moregenans}) exceeds in an essential way a number of
extra coef\/f\/icients $a_k$,  $b_r$ to be determined.

So putting aside  attempts to discover the analytical expression
for soliton-like solutions of the GBE,  we are searching for
approximated solutions to the equation (\ref{redode}), represented
by the series (\ref{approduf}),
 \begin{gather}\label{approduf}
U(T)= \left\{
\begin{array}{ll}
\displaystyle \sum_{n=1}^\infty a_n\exp \left[\alpha  T n \right] & \quad \mbox{when} \quad  T<0, \vspace{2mm}\\
\displaystyle \sum_{n=1}^\infty b_n\exp \left[-\beta  T n \right]
& \quad \mbox{when} \quad T>0,
\end{array}
\right.
\end{gather}
with the additional condition
\begin{gather}\label{initduf}
U(0)=x_{*}, \qquad U'(0)=0,
\end{gather}
where $x_{*}$ is a  point of intersection of the homoclinic
trajectory with the horizontal axis, which, generally speaking,
can be obtained from the numerical experiment. With this
conditions the two branches of the homoclinic trajectory are
``tied up'' in the point $(x_*,  0)$.

There are two ways of determining the unknown constants $a_k$ and
$\alpha$. Firstly, it is possible to insert the series
\begin{gather}\label{seriesup}
\varphi (T )=\sum_{k=1}^\infty a_k\exp \left[\alpha  T k \right]
\end{gather}
into (\ref{genburg}) and equate to zero the coef\/f\/icients
standing at dif\/ferent powers of  the function $\exp{
\left[\alpha
 T\right] }.$ This way we obtain the recurrence $a_{k+1}=\psi
_{k+1}(a_1,  a_2,\dots, a_k)$. Another possibility is to use the
initial conditions (\ref{initduf}) and their dif\/ferential
consequences. As in the previous case, this approach is quite
algorithmic for any second order ODE that is linear with respect
to the highest derivatives. The approach is based on the following
statement,  which can be easily proved by means of mathematical
induction.

\begin{lemma}\label{lem:4-2}
Consider the Cauchy problem
\begin{gather}\label{absteq}
U''(\xi)=F(\xi, U, U'),
\\
\label{abstini} U(0)=x_0, \qquad U'(0)=x_1.
\end{gather}
Suppose we are looking for the approximate solution to
\eqref{absteq}, \eqref{abstini} in the form
\begin{gather}\label{abstser}
U(\xi)=\sum_{k=1}^N a_k\exp \left[\alpha \xi k \right].
\end{gather}
Then the coefficients $\left\{a_k\right\}_{k=1}^N$ can be obtained
from the conditions \eqref{abstini} and the $N-2$ equations
\[
U^{(j+2)}(0)=\sum_{k=1}^N (\alpha  k)^{j+2} a_k= D^{j}_\xi F(\xi,
U, U')|_{\xi=0}=\Phi_j\big(U(0), U'(0),\dots , U^{(j+1)}(0)\big),
\]
 $j=0, 1,\dots, N-3$, where $D_\xi$  denotes the total derivative. The functions
$\Phi_j(U(0),\dots, U^{(j+1)}(0))$   are recurrently expressed
merely by the initial  conditions $x_0$,  $x_1$ and do not depend
upon $a_k$.
\end{lemma}

So according to Lemma~\ref{lem:4-2}, using Cauchy data
(\ref{abstini}) and their dif\/ferential consequences we get the
linear system
\begin{gather}\label{linsys}
\left[
\begin{array}{ccccc}
(\alpha\cdot 1)^0 & (\alpha\cdot 2)^0 & (\alpha\cdot 3)^0 & \cdots  & (\alpha\cdot N)^0 \\
(\alpha\cdot 1)^1 & (\alpha\cdot 2)^1 & (\alpha\cdot 3)^1 & \cdots  & (\alpha\cdot N)^1 \\
(\alpha\cdot 1)^2 & (\alpha\cdot 2)^2 & (\alpha\cdot 3)^2 & \cdots  & (\alpha\cdot N)^2 \\
\cdots & \cdots  & \cdots  & \cdots  & \cdots      \\
(\alpha\cdot 1)^{N-1} & (\alpha\cdot 2)^{N-1} & (\alpha\cdot
3)^{N-1} & \cdots  & (\alpha\cdot N)^{N-1}
\end{array}
\right] \left[
\begin{array}{c}
a_1 \\
a_2 \\
a_3 \\
\cdots  \\
a_N
\end{array}
\right]= \left[
\begin{array}{c}
x_0 \\
x_1 \\
\Phi_0 \\
\cdots  \\
\Phi_{N-3}
\end{array}
\right].
\end{gather}
The matrix $\hat M_N$ standing at the RHS of equation
(\ref{linsys}) is of the Vandermonde type \cite{korn}. Its
determinant is given by the formula
\[
\det{\hat M_N}=\alpha^{\frac{N (N-1)}{2}}\cdot 0! \cdot 1! \cdot
2! \cdots  (N-1)!,
\]
so it is always nonsingular and the system has a unique solution
for each $N$.

\begin{remark*}
 The above lemma says nothing about the
convergence of the series (\ref{abstser}) to the real solution of
the initial value problem (\ref{absteq}), (\ref{abstini}) as $N\to
\infty$. This problem needs a special investigation.
\end{remark*}

 Our analysis suggests that the upper and
lower branches of the homoclinic trajectory are so strongly
asymmetric that the convergency of the approximated solution based
on the f\/irst method of determining the coef\/f\/icients of
series expansions (\ref{approduf}) takes place merely for the
lower one, whereas the upper branch should be approximated with
the second method only.

\begin{figure}[t]
\begin{minipage}[t]{7.5cm}
\centerline{\includegraphics[width=7cm]{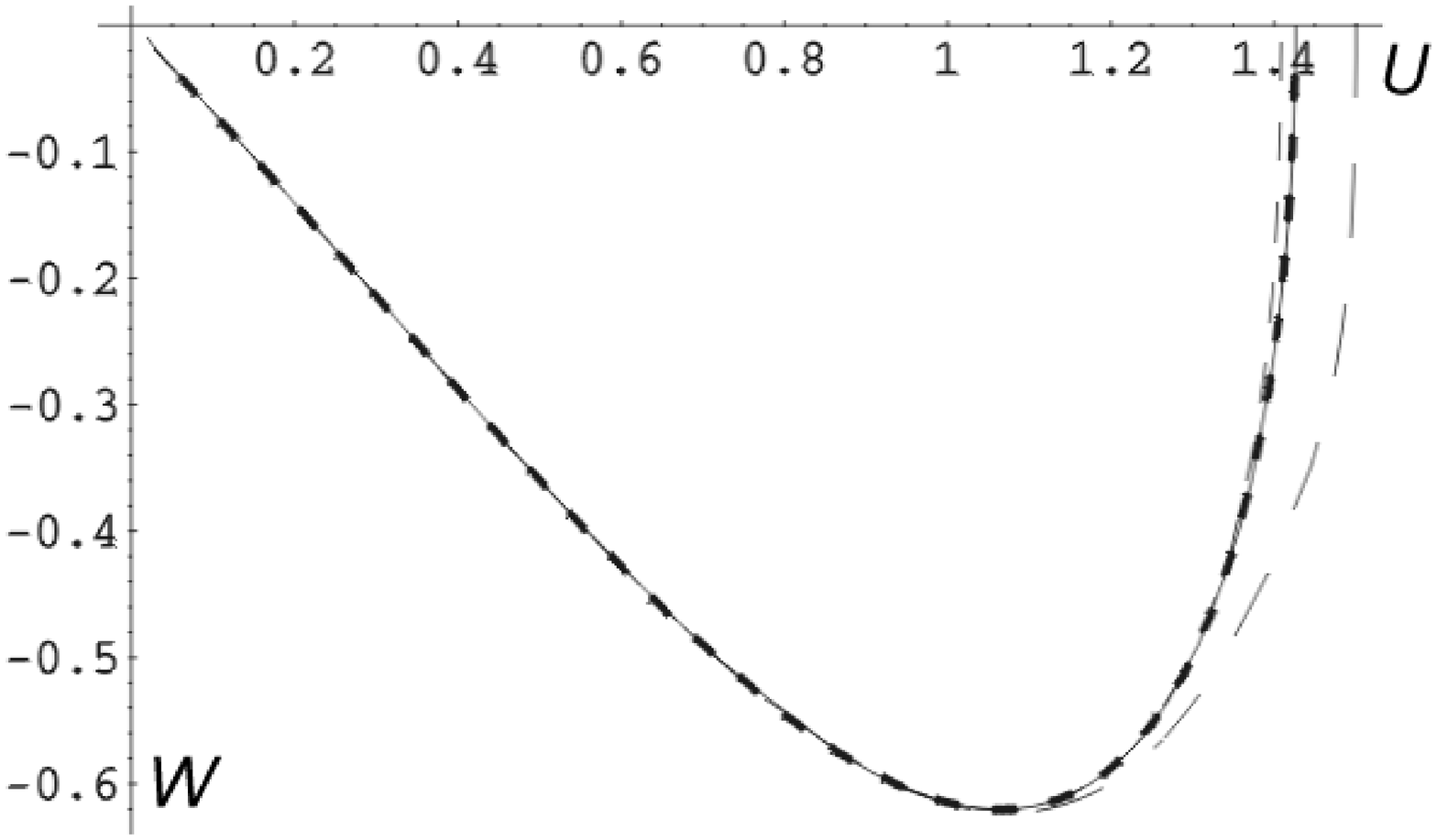}}
\vspace{-2mm} \caption{The lower branch of the homoclinic loop  of
system (\ref{candis}): numerical solution (solid) and the
truncated series (\ref{approduf}) (dashed). The coef\/f\/icients
$a_k$ are def\/ined by means of the f\/irst method. The bold
dashed line is obtained for $N=40$.}\label{fig:4-2}
\end{minipage}
\hfill
\begin{minipage}[t]{7.5cm}
\centerline{\includegraphics[width=7.5cm]{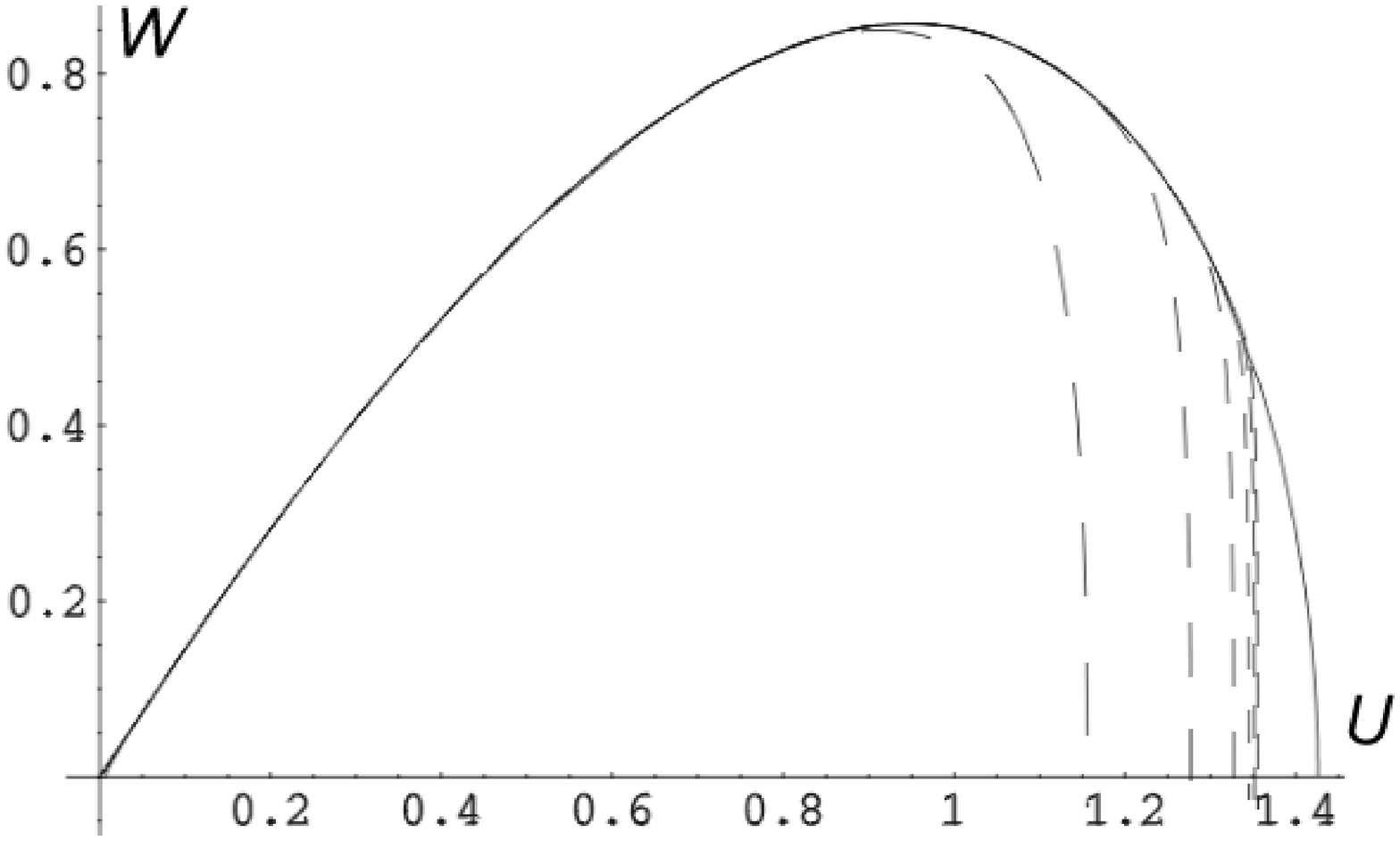}}
\vspace{-2mm} \caption{The upper branch of the homoclinic loop  of
system (\ref{candis}): numerical solution (solid) and the
truncated series (\ref{approduf}) (dashed). The coef\/f\/icients
$a_k$ are def\/ined by means of the f\/irst
method.}\label{fig:4-3}
\end{minipage}
\vspace{-1mm}
\end{figure}

Thus, we approximate the lower branch of homoclinic solution by
the series
\begin{gather*}
\psi (T )=\sum_{n=1}^\infty b_n\exp \left[-\beta  T n \right],
\end{gather*}
truncated on a number $N.$  The f\/inite series is inserted into
the factorized equation (\ref{redode}) and the coef\/f\/icients
standing at the subsequent powers of function $\exp[-\beta T]$ are
equated to zero. For those values of the parameters that have been
employed in the Section~2, the coef\/f\/icient at
$\left\{\exp[-\beta T]\right\}^1$ nullif\/ies if $\beta\approx
0.6658$. Equating to zero the coef\/f\/icients of
$\left\{\exp[-\beta T]\right\}^k,$ we obtain  the recurrence for
the coef\/f\/icients $b_k$, $k=2,3,\dots ,N$, while $b_1$ remains
unidentif\/ied. We omit the presentation of this recurrence since
it is irregular and cannot be expressed it in the analytical form.
Therefore all the calculations were performed with the help of the
``Mathematica'' package. They were based in an essential way upon
the assumption that the phase trajectory approaches the point
$U=x_{*}=1.426095$ of the horizontal axis as $T\to 0$. Approximate
solutions corresponding to dif\/ferent values of the number $N$
are shown in Fig.~\ref{fig:4-2}.

\begin{figure}[t]
\begin{minipage}[t]{7.5cm}
\centerline{\includegraphics[width=7.5cm]{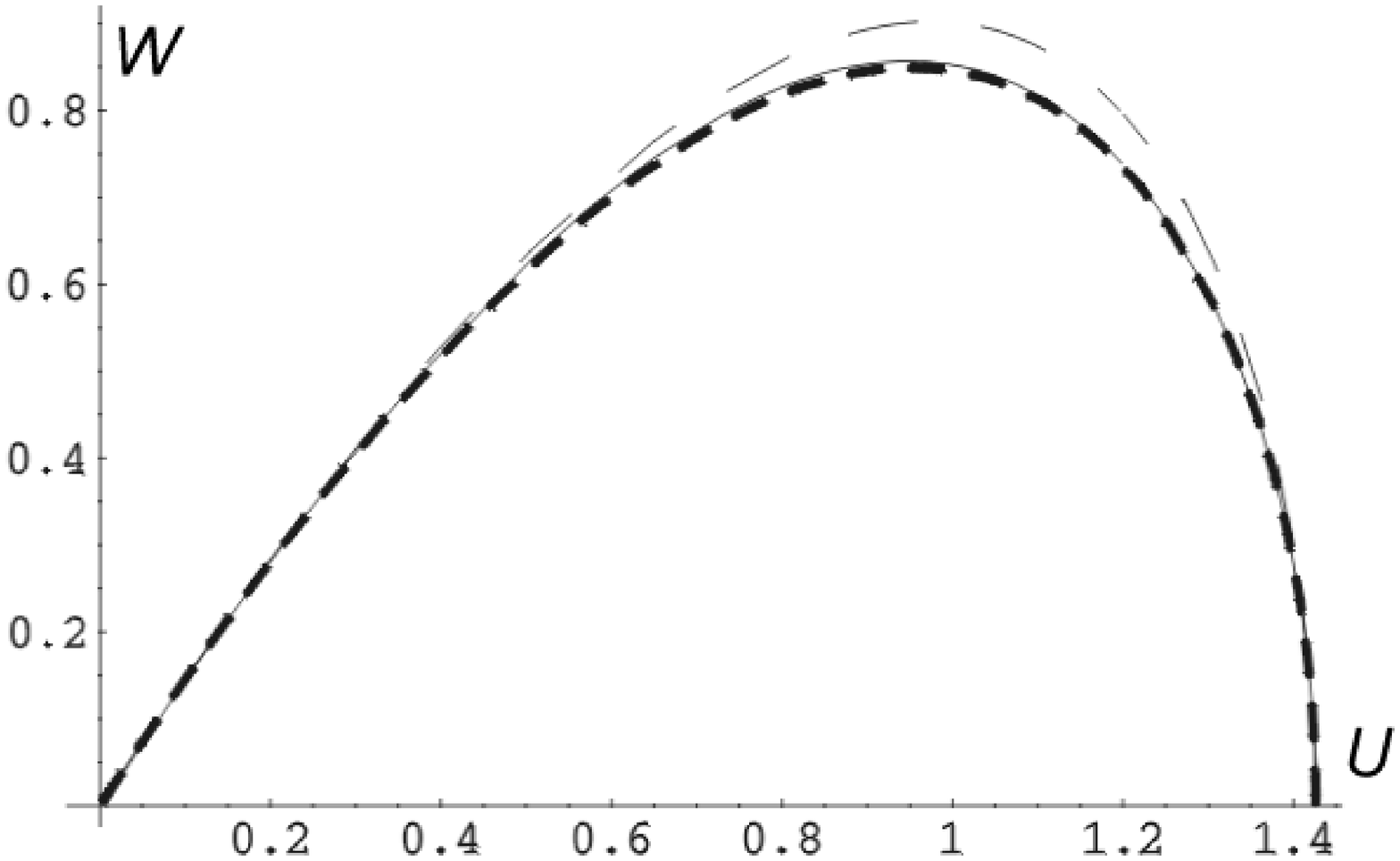}}
\vspace{-2mm} \caption{The upper branch of the homoclinic loop  of
system (\ref{candis}): numerical solution (solid) and the
truncated series (\ref{approduf}) (dashed). The coef\/f\/icients
$a_k$ are def\/ined by means of the second method. The bold dashed
line corresponds to $N=20$.}\label{fig:4-4}
\end{minipage}
\hfill
\begin{minipage}[t]{7.5cm}
\centerline{\includegraphics[width=7.5cm]{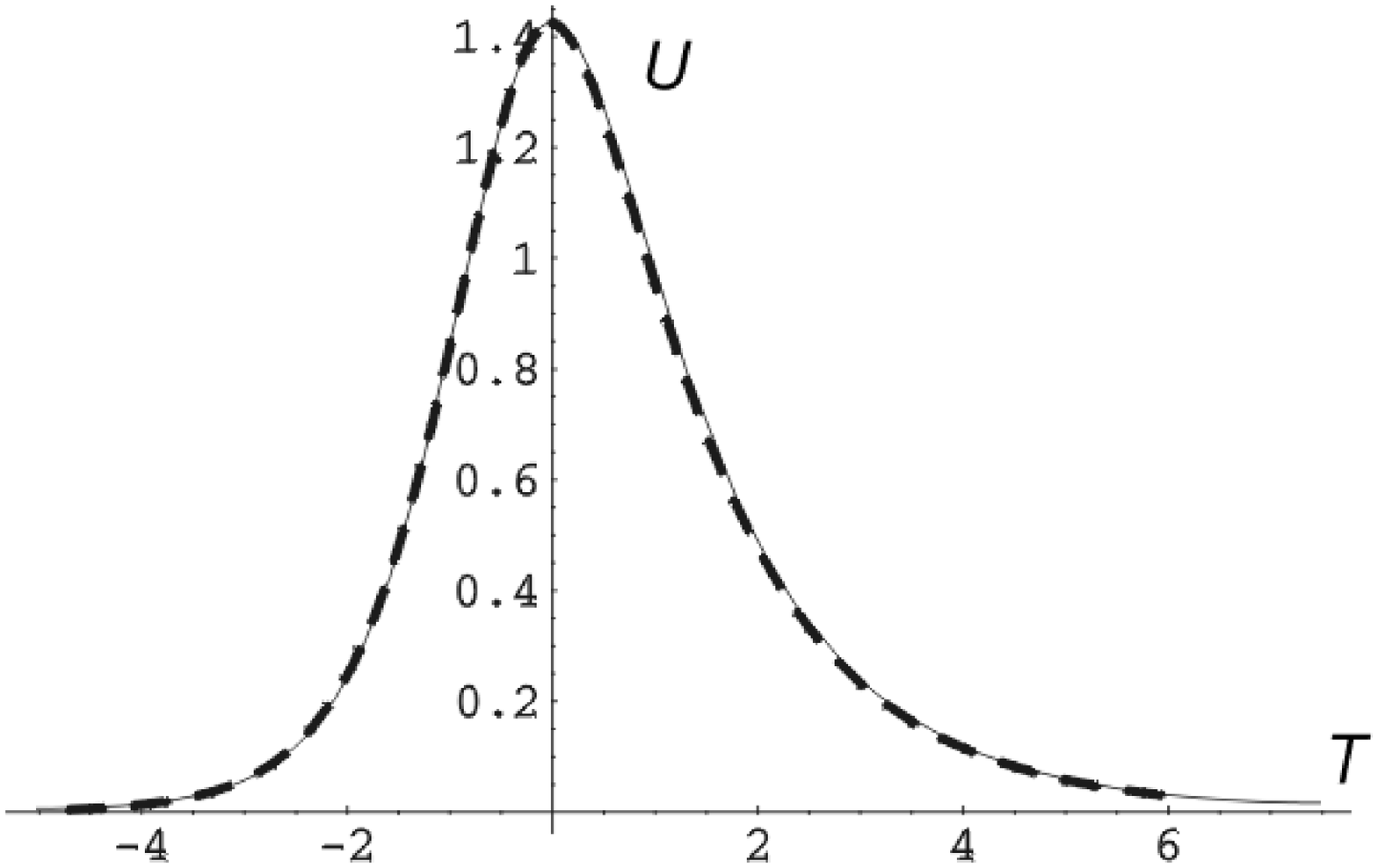}}
\vspace{-2mm} \caption{Solitary wave solution of equation
(\ref{genburg}) (solid) and its approximation
(dashed).}\label{fig:4-5}
\end{minipage}
\vspace{-4mm}
\end{figure}

Unfortunately we fail to employ a similar series decomposition and
the method of f\/inding the unknown coef\/f\/icients for the upper
branch, since the series do not converge to the numerical solution
in proximity of the far end of the homoclinic trajectory, as it is
seen in Fig.~\ref{fig:4-3}. In fact, the series is convergent yet
the limiting function does not coincide with the solution,
obtained by means of the highly precise numerical scheme. So in
order to determine the series coef\/f\/icients, we turn to the
second scheme. Keeping in the series (\ref{seriesup}) $N$ elements
and substituting the truncated series into equation
(\ref{redode}), we obtain from (\ref{initduf}) two conditions for
the coef\/f\/icients $a_k$:
\begin{gather}\label{uzero}
U(0)=\sum_{k=1}^{N} a_k=x_* = 1.426095, \qquad
U'(0)=\sum_{k=1}^{N}k a_k=0.
\end{gather}
The f\/irst dif\/ferential consequence of these  conditions gives
us equation (\ref{redode}) itself. Taking it into the point $T=0$
and employing equations (\ref{uzero}), we obtain:
\begin{gather*}
U''(0)=\sum_{k=1}^{N}(\alpha k)^2 a_k=(1-x_*^2) x_*=f_3(x_*).
\end{gather*}
In order to def\/ine all the coef\/f\/icients of the series
truncated on $N$-th element we need $N-3$ extra conditions. We can
obtain these missing conditions from further  dif\/ferential
consequences, as it is stated in Lemma~\ref{lem:4-2}.
Coef\/f\/icients $\left\{a_k\right\}_{k=1}^N$ satisfy the linear
system (\ref{linsys}) that delivers a~unique solution, providing
the RHS and the parameter $\alpha$ are determined. The remaining
$N-3$ components of the vector, positioned at the RHS of system
(\ref{linsys}), were obtained by means of some recurrent
procedure, written in ``Mathematica'' package.
 In order to obtain the parameter~$\alpha$, we
insert the solution (\ref{seriesup}) into (\ref{redode}) and
equate to zero the coef\/f\/icient $ \alpha^2+A \mu \alpha-1, $ at
the lowest power of the exponential function. For $A=1$ and
$\mu=-0.836$ this gives us the value $\alpha=1.5018$.

The above procedure was used to obtain the upper branch of the
homoclinic curve of system~(\ref{candis}) (or what is the same,
the left ``wing'' of the SW solution of the equation
(\ref{redode})). The results obtained are shown in
Fig.~\ref{fig:4-4}. The upper and lower branches sewed together
are shown in Fig.~\ref{fig:4-5}, from which we can see that
suggested approach  quite well approximates the homoclinic
solution of equation (\ref{genburg}).

\section{Concluding remarks}

We have reviewed in this study the methods and tools that allow
one to describe the solutions of the nonlinear PDEs with the
required features and asymptotic behavior. The exact solutions
presented here are obtained with the help of the ansatz
(\ref{expans})\footnote{During the preparation of this proof, we
wre pointed to the paper \cite{cornille} in which similar ansatzes
were applied to a family of weakly nonlinear equations of
Boltzmann type.}, which proves to be ef\/fective when the methods
described in \cite{fan,nikbar} fail. We concentrated mainly upon
the soliton-like TW solutions and their description, yet similar
tools can be applied to the description of  kink-like, periodic
and multi-periodic solutions as well.

In contrast to the classical self-similarity methods, ansatz-based
methods \cite{fan,nikbar,baryur,vladku} enable to obtain the exact
solution with required properties an asymptotics (such e.g.\ as
the periodic, soliton-like and kink-like TW solutions). This is
due to the fact that within these methods solutions are def\/ined
as algebraic combinations of certain set of functions (elementary
of special ones), closed with respect to dif\/ferentiation and
algebraic operations, being represented in a~form, enabling to
control some features and the asymptotic behavior of the solutions
to be described. Yet efficient employment of the direct algebraic
balance methods is related to  one's ability to solve large
systems of nonlinear algebraic equations. In addition to this
purely technical obstacle, there is a number of evidence that none
of the already known version of the balance algebraic method is
fully universal.
 For this reason the authors put forward the method of the asymptotic
description of the soliton-like TW solutions by an inf\/inite
series of exponential functions. The ef\/fectiveness of this
method, connected with some general properties of the set of
exponential functions \cite{leontiev}, is demonstrated by its
application to the GBE. Previous qualitative investigations showed
the existence of such solutions for the specif\/ic values of the
parameters and further on it was shown that these solutions cannot
be described as f\/inite algebraic combination of exponential
functions.

Let us say a few words about the problem   of the truncated series
(\ref{approduf}) convergence to the exact solution.  A uniform
convergency of a series of exponential functions to an exact
solution can be easily stated e.g.\ for the nonlinear wave
equation, because  the expressions for the coef\/f\/icients~$a_n$
and $b_n$ are known in this case~\cite{romp05}. Unfortunately, for
GBE the recurrent expressions for indices~$a_n$,   $b_n$ of the
decomposition (\ref{approduf}) are not known and the problem of
the series convergence remains still open.

The method of approximation of solitary wave regimes, described in
Section~3, is relatively easy to perform in case when the
factorized system is two-dimensional. In the less trivial cases,
when the factorized system is multidimensional, the homoclinic
trajectories represen\-ting the soliton-like solutions become much
more complicated. As an example let us mention a~Shil'nikov
homoclinic trajectory associated with the saddle-focus of a
three-dimensional dynamical system~\cite{Shiln,G-H}. In order to
deliver an approximated description of such trajectory following
the methodology outlined in Section~3 of this study, it is
necessary to incorporate wider class of functions. In earlier
attempts at describing the Shil'nikov homoclinics \cite{shkadov},
a combination of the exponential and trigonometric functions was
used. The approximation employed in \cite{shkadov} was not very
ef\/fective because of the great technical dif\/f\/iculties
encountered. Since then a number of powerful programs for the
symbolic calculations were developed, and it is our hope that
approximation similar to that used in  \cite{shkadov} actually can
be performed technically much more ef\/fectively.

\subsection*{Acknowledgments} The authors thank the anonymous referees
for helpful suggestions concerning the presentation of results.

\LastPageEnding

\end{document}